\documentclass[aps,pra,preprint,groupedaddress,showpacs]{revtex4}

\begin{document}

\title{Application of the no-signaling principle to obtain quantum
cloners for any allowed value of fidelity}

\author{Z. Gedik}

\email[]{gedik@sabanciuniv.edu}

\author{B. \c{C}akmak}

\affiliation{Faculty of Engineering and Natural Sciences, Sabanci
University, 34956 Tuzla, Istanbul, Turkey}

\date{\today}

\begin{abstract}
Special relativity forbids superluminal influences. Using only the
no-signaling principle and an assumption about the form of the
Schmidt decomposition, we show that for \emph{any} allowed fidelity
there is a \emph{unique} approximate qubit cloner which can be
written explicitly. We introduce the prime cloners whose fidelities
have multiplicative property and show that the fidelity of the prime
cloners for the infinite copy limit is 1/2.
\end{abstract}


\maketitle

\section{Introduction}

According to theory of special relativity, it is not possible to
send instantaneous signals between two spatially separated
observers. The no-signaling (NS) principle is necessary for
consistency of the theory of relativity and quantum mechanics. In
this work, we show that NS principle can be used to \emph{derive}
universal and symmetric 1-to-$M$ qubit cloning transformation for
\emph{any} allowed value of fidelity.

Impossibility of faster than light communication based on quantum
correlations, after presentation of a signaling protocol based on
perfect cloning~\cite{Herbert}, led to the discovery of the
no-cloning theorem~\cite{Wootters-Zurek,Dieks}. However, the NS
principle leaves room for an approximate cloning. Imperfect or
approximate 1-to-2 optimal quantum cloners have been shown to
exist~\cite{Buzek-Hillery,Bruss1} and the results have been
generalized to 1-to-$M$ cloning~\cite{Gisin-Massar}. An expression
for the maximum fidelity of $N$-to-$M$ cloning of qudits has been
found~\cite{Werner}, and corresponding cloners have been
obtained~\cite{Fan,Scarani,FanR} . In this work, we obtain the
universal symmetric quantum cloners for any allowed fidelity,
including the best (optimal) one, using the NS principle. We show
that fidelity determines the cloning transformation uniquely.

Gisin has analyzed universal symmetric 1-to-2 cloning under the NS
principle and has shown that the optimal value is the same as that
of the optimal quantum cloner~\cite{Gisin}. It has been argued that
modification of the quantum theory by introducing nonlinear time
evolution for pure states \cite{Weinberg1,Weinberg2} might lead to
superluminal communication \cite{Gisin1,Czachor,Polchinski}, and
hence, the NS principle implies linearity of quantum mechanics.
Inspired by the Gisin's formalism, Simon has used the
\emph{linearity} of quantum dynamics to rederive the \emph{optimal}
1-to-$M$ cloners~\cite{SimonT}. In this work, we apply \emph{NS
principle directly}, rather than utilizing the linearity, to
1-to-$M$ cloning, and we obtain the unique cloners for \emph{all
possible} values of fidelity. Furthermore, we construct the prime
cloners which have the property that the fidelity of a 1-to-$MN$
cloner can be obtained by the successive use of 1-to-$M$ and
1-to-$N$ cloners. For a given number of copies, we obtain the
\emph{unique} prime cloner.

The article is organized as follows. We first introduce the
pseudo-spin formalism for universal symmetric cloning. Next, we
discuss the consequences of impossibility of instantaneous
signaling. We examine the implications of the NS principle on the
cloning transformation, and hence, explicitly obtain all possible
values of fidelity along with the corresponding cloners. We then
derive the quantum cloners along with what we call \emph{prime
cloners}.

\section{Pseudo-spin Formalism}

Symmetry of the output state, namely invariance of the wave function
under the exchange operation, reduces the dimension of the Hilbert
space from $2^M$ to $M+1$. Pseudo-spin formalism utilizes this
dimensional reduction. Let $|\hat{n}\rangle$ be the state vector of
the qubit to be cloned. In the so-called pseudo-spin representation,
we treat the qubit as a spin-1/2 object, and thus $|\hat{n}\rangle$
corresponds to a spin-up state in the $\hat{n}-$direction. Then,
symmetric $M-$qubit states can simply be represented by the total
spin states with $j=M/2$. Therefore, we can use the states
\begin{equation}\label{ps}
|\hat{n};jm\rangle=\left(\begin{array}{c}
                                                         2j \\
                                                         j+m \\
                                                       \end{array}
                                                     \right)^{-1/2}
\mathcal{P}\{\underbrace{|\hat{n}\rangle\otimes...\otimes|\hat{n}\rangle}_{j+m}\otimes\underbrace{|-\hat{n}\rangle\otimes...\otimes|-\hat{n}\rangle}_{j-m}\}
\end{equation}
as the basis elements in the  $M-$qubit symmetric space. Here,
$\mathcal{P}$ denotes all possible permutations of the product state
in the parentheses and  $m=-j,-j+1,...,j-1,j$. Pseudo-spin
formulation allows us to solve the problem by using the techniques
of rotations in quantum mechanics.

Any quantum operation performed on qubits can be modeled as a
unitary operation acting on the qubits plus an ancillary system. In
the case of cloning, this system is called the cloning machine.
After the cloning interaction, the $M$~qubits will in general be
entangled with the cloning machine, and the state of the whole
system will be pure. This pure entangled state can be written in the
Schmidt form. We assume that the Schmidt basis for $M$~qubits
consists of the states $|\hat{n};jm\rangle$. Therefore, in the most
general sense, the transformation for universal and symmetric pure
state cloning is given by
\begin{equation}\label{trans}
|\hat{n}\rangle\otimes\underbrace{|0\rangle\otimes...\otimes|0\rangle}_{M-1}\otimes|R\rangle\rightarrow\sum_{m=-j}^{j}a_{jm}|\hat{n};jm\rangle\otimes|R_{jm}\left(\hat{n}\right)\rangle,
\end{equation}
where $|0\rangle$ and $|R\rangle$ are blank copy and initial machine
states, respectively. The normalization of the output state implies
that $\sum_m p_{jm}=1$, where $p_{jm}=a_{jm}^2$. Independence of the
probabilities $p_{jm}$ from $\hat{n}$ is necessary for the
transformation to be universal. As a result of the Schmidt
decomposition, the machine states
$|R_{jm}\left(\hat{n}\right)\rangle$, are orthonormal, i.e.,
$\langle
R_{jm}\left(\hat{n}\right)|R_{jm'}\left(\hat{n}\right)\rangle=\delta_{mm'}$.
After tracing out the states of the machine, we can formulate the
problem in terms of the original state and its copies. The reduced
transformation becomes
\begin{equation}\label{redtran}
T_j\left(|\hat{n}\rangle\langle\hat{n}|\right)=\sum_{m=-j}^{j}p_{jm}|\hat{n};jm\rangle\langle\hat{n};jm|.
\end{equation}
We see that due to the orthonormality of the machine states, the
output state of the cloning transformation is described by a
diagonal density matrix.

Fidelity, the measure of the quality of cloning, is defined as the
projection of the final single qubit state (obtained by tracing out
the other $M-1$ qubits) onto the original state. Therefore, $(j-m)$
combinations of $2j-1$ elements of the sum in (\ref{redtran})
contribute to the fidelity  expression, and the resulting value is
given by
\begin{equation}\label{fid}
F_j=\sum_{m=-j}^{j}\left(\begin{array}{c}
                                                         2j-1 \\
                                                         j-m \\
                                                       \end{array}
                                                     \right)\left(
                                                       \begin{array}{c}
                                                         2j \\
                                                         j-m \\
                                                       \end{array}
                                                     \right)^{-1}p_{jm}=\frac{1}{2}\left(1+\frac{1}{j}\sum_{m=-j}^{j}mp_{jm}\right).
\end{equation}
Fidelity is a linear function of the expectation value of the
$z-$component of the pseudo-angular momentum. If perfect cloning
were possible, we would have $p_{jm}=\delta_{mj}$, which results in
$F_j=1$. In the next section, we shall evaluate the upper and lower
limits for $F_j$ when the NS principle is taken as a constraint.

\section{No-signalling Constraint}

The impossibility of superluminal communication implies that
transforms of indistinguishable mixtures are also indistinguishable.
This is because, two observers can share entangled states, where one
of them can perform projective measurements to determine the
spectral decomposition of the reduced density matrix of the other
observer~\cite{Schrodinger,Jaynes,Hadjisavvas,Gisin2,Hughston,Mermin,Kirkpatrick}.
Hence, for a given transformation $f$, when two convex linear
combinations are equal to
$\sum_ix_i|\psi_i\rangle\langle\psi_i|=\sum_jy_j|\phi_j\rangle\langle\phi_j|$,
so too must their images, i.e.,
$\sum_ix_if\left(|\psi_i\rangle\langle\psi_i|\right)=\sum_jy_jf\left(|\phi_j\rangle\langle\phi_j|\right)$,
to prevent signaling. We note that this is a condition involving
maps of pure states only, and it implies that
$\sum_ix_if\left(|\psi_i\rangle\langle\psi_i|\right)$ should be a
function of only $\sum_ix_i|\psi_i\rangle\langle\psi_i|$. Therefore,
the NS principle requires that
\begin{equation}\label{ns1}
\sum_ix_if\left(|\psi_i\rangle\langle\psi_i|\right)=g\left(\sum_ix_i|\psi_i\rangle\langle\psi_i|\right),
\end{equation}
where the map $g$ is not necessarily same as $f$. Equivalence of $f$
and $g$ cannot be concluded from the above form of the NS principle.
However, by introducing a proper communication protocol, we can show
that $f=g$, and thus, convex linearity is a consequence of the NS
condition~\cite{Gedik}. Now, since
$\left(|\hat{n}\rangle\langle\hat{n}|+|-\hat{n}\rangle\langle-\hat{n}|\right)/2$
is equivalent to the identity operator for any $\hat{n}$, all its
images, i.e., $M$ clones, should be invariant under changes in
$\hat{n}$, too. Since $|-\hat{n};jm\rangle=|\hat{n};j,-m\rangle$,
the indistinguishability requirement states that
\begin{equation}\label{nosign}
T_j\left(|\hat{n}\rangle\langle\hat{n}|\right)+T_j\left(|-\hat{n}\rangle\langle-\hat{n}|\right)=\sum_{m=-j}^{j}\left(p_{jm}+p_{j,-m}\right)|\hat{n};jm\rangle\langle\hat{n};jm|
\end{equation}
is rotationally invariant in the pseudo-spin space, and thus, the
coefficients of expansion  should be independent of $m$, i.e.,
\begin{equation}\label{nosignc}
p_{jm}+p_{j,-m}=\frac{2}{2j+1}.
\end{equation}
Equation~(\ref{nosignc}) is satisfied by any universal symmetric
cloner. However, NS principle is more restrictive than the
constraint given by rotational invariance of the expression given in
(\ref{nosign}). Let us consider two arbitrary qubit states
$|\hat{n}\rangle$ and $|\hat{n}'\rangle$, and their arbitrary convex
linear combination
$\rho=r|\hat{n}\rangle\langle\hat{n}|+(1-r)|\hat{n}'\rangle\langle\hat{n}'|$
where $0\leq r\leq1$. The density matrix $\rho$ is diagonal for some
$|\hat{m}\rangle$, and hence, it can be written as
$\rho=s|\hat{m}\rangle\langle\hat{m}|+(1-s)|-\hat{m}\rangle\langle-\hat{m}|$
with $0\leq s \leq 1$. Different convex decompositions of the
density matrix $\rho$ can be obtained by different choices of
discrete measurements performed by another observer sharing an
entangled state with the first observer. In order to prevent
signaling, these two representations of the same density matrix must
have the same images under the transformation. Therefore,
\begin{equation}\label{lin}
rT_j\left(|\hat{n}\rangle\langle\hat{n}|\right)+(1-r)T_j\left(|\hat{n}'\rangle\langle\hat{n}'|\right)=sT_j\left(|\hat{m}\rangle\langle\hat{m}|\right)+(1-s)T_j\left(|-\hat{m}\rangle\langle-\hat{m}|\right).
\end{equation}
We can choose our coordinate axes so that $\hat{m}=\hat{z}$. Then,
$s|\hat{m}\rangle\langle\hat{m}|+(1-s)|-\hat{m}\rangle\langle-\hat{m}|$
becomes
\begin{equation}
\frac{1}{2}\left(1+\frac{\sin(\theta+\theta')}{\sin\theta+\sin\theta'}\right)|\hat{z}\rangle\langle\hat{z}|+\frac{1}{2}\left(1-\frac{\sin(\theta+\theta')}{\sin\theta+\sin\theta'}\right)|-\hat{z}\rangle\langle-\hat{z}|,
\end{equation}
where $\theta$ ($\theta'$) is the angle between $\hat{z}$ and
$\hat{n}$ ($\hat{n}'$), and
$r=\sin\theta'/(\sin\theta+\sin\theta')$. Therefore, the NS
constraint takes the form
\begin{equation}\label{lincom}
\sum_{m=-j}^{j}\left(c_+p_{jm}+c_-p_{j,-m}\right)|\hat{z};jm\rangle\langle\hat{z};jm|=\sum_{m=-j}^{j}p_{jm}\left(\sin\theta'|\hat{n};jm\rangle\langle\hat{n};jm|+\sin\theta|\hat{n}';jm\rangle\langle\hat{n}';jm|\right),
\end{equation}
where $2c_\pm=\sin\theta+\sin\theta'\pm\sin(\theta+\theta')$. We
note that
$|\langle\hat{z};jm|\hat{n};jm'\rangle|=|d_{mm'}^{(j)}(\theta)|$,
where $d_{mm'}^{(j)}(\theta)$ are the elements of the reduced Wigner
rotation matrix. Similarly, we have
$|\langle\hat{z};jm|\hat{n}';jm'\rangle|=|d_{mm'}^{(j)}(\theta')|$.
Therefore, (\ref{lincom}) can be written as
\begin{equation}
c_+p_{jm}+c_-p_{j,-m}=\sum_{m'=-j}^{j}\left(|d_{mm'}^{(j)}(\theta)|^2
\sin\theta'+|d_{mm'}^{(j)}(\theta')|^2 \sin\theta\right)p_{jm'}.
\end{equation}
Finally, using the constraint (\ref{nosignc}), the NS principle can
be written as an eigenvalue equation
\begin{equation}\label{eig}
\sum_{m'=-j}^{j}\left(|d_{mm'}^{(j)}(\theta)|^2
\sin\theta'+|d_{mm'}^{(j)}(\theta')|^2
\sin\theta-\frac{2c_-}{2j+1}\right)p_{jm'}=\sin(\theta+\theta')p_{jm}.
\end{equation}
Since $|d_{mm'}^{(j)}(\theta)|=|d_{-m,-m'}^{(j)}(\theta)|$ when
$p_{jm}$ is a solution of (\ref{eig}), $p_{j,-m}$ is also a solution
with the same eigenvalue. In other words, eigenvectors are (or, in
case of degeneracy, can be chosen to be) either symmetric (even) or
anti-symmetric (odd) in $m$.

Let us assume that $p_{jm}$ can be written as an analytic function
$f(m)$ of $m$. Since
$|d_{mm'}^{(j)}(\theta)|=|d_{m'm}^{(j)}(\theta)|$, we have
\begin{equation}
\sum_{m'=-j}^{j}|d_{mm'}^{(j)}(\theta)|^2f(m')=\sum_{m'=-j}^{j}\langle\hat{n};jm'|f(J_z)|\hat{z};jm\rangle\langle\hat{z};jm|\hat{n};jm'\rangle=f(m\cos\theta).
\end{equation}
We see that $\sin(\theta+\theta')$ is a two-fold degenerate
eigenvalue, and $p_{jm}=1/(2j+1)$ is the only symmetric solution
whereas $p_{jm}=\pm m/j(2j+1)$ are the only possible anti-symmetric
solutions. The positivity of $p_{jm}$'s allows us to write two
linearly independent solutions as $(j+m)/j(2j+1)$ and
$(j-m)/j(2j+1)$. Hence, the most general solution becomes
\begin{equation}\label{pm}
p_{jm}(t)=t\frac{j+m}{j(2j+1)}+(1-t)\frac{j-m}{j(2j+1)},
\end{equation}
where $0\leq t\leq 1$. The corresponding fidelity is given by
\begin{equation}\label{fidq}
F_j(t)=\frac{2j-1+2(j+1)t}{6j},
\end{equation}
which has its maximum value at $(4j+1)/6j$ when $t=1$. This is the
well known optimal quantum cloner fidelity~\cite{Gisin-Massar}. In
this case, $p_{jm}$ coefficients become identical to the optimal
quantum machine coefficients. We observe that $p_{j,-j}$ vanishes
only for the optimal cloner. Therefore, if we exclude the worst
cloning case from the set of possible output states by assuming that
$p_{j,-j}=0$ (as has been done in Refs.~\cite{Gisin-Massar} and
\cite{Fan}), we cannot find the universal cloners other than the
optimal one. That is, the optimal cloner is the only universal
quantum cloning machine for which the state $|\hat{n};j,-j\rangle$
has zero probability.

Equations (\ref{pm}) and (\ref{fidq}) can be used to find the
quantum cloner for a given fidelity $F_j$ in the allowed interval
$\left[1-\left(F_j\right)_{\text{max}},
\left(F_j\right)_{\text{max}}\right]$. They can also be used to
construct a quantum cloner satisfying some specific property. For
example, let us consider the cloners where successive use of
1-to-$M$ and 1-to-$N$ cloners gives the same fidelity as a single
1-to-$MN$ cloner. We call such a cloner as \emph{prime cloner} since
it is enough to have 1-to-$p$ cloners, where $p$ is a prime number,
to construct any 1-to-$M$ cloner. The fidelities of prime cloners
$F^P$ should satisfy
\begin{equation}\label{pclon}
F^{\text{P}}_{M/2}F^{\text{P}}_{N/2}+\left(1-F^{\text{P}}_{M/2}\right)\left(1-F^{\text{P}}_{N/2}\right)=F^{\text{P}}_{MN/2}.
\end{equation}
Substituting the fidelity expressions given by (\ref{fidq}), we find
the coefficients of expansion as
\begin{equation}\label{pclonc}
p_{jm}=\frac{1}{2j+1}\left(1+\frac{3m}{2j(j+1)}\right)
\end{equation}
which corresponds to $t=(2j+5)/4(j+1)$. We note that fidelity
$F^{\text{P}}_j=(2j+1)/4j$ tends to $F^{\text{P}}_\infty=1/2$ which
is just at the center of the allowed fidelity interval at infinite
copy limit.

\section{Conclusion}

We presented a method for constructing universal symmetric 1-to-$M$
qubit cloners. In particular, we systematically derived the
properties of universal symmetric quantum cloning machines instead
of postulating them first and proving them afterwards. Direct use of
NS principle allowed us to find the best (optimal) and the worst
cloners along with all other cloners having a fidelity between the
maximum and the minimum values. For a given fidelity, cloning
transformation is unique. We introduced the prime cloners whose
fidelities have multiplicative property and we found the
corresponding machines.

\section{Acknowledgement}

This work has been partially supported by the Scientific and
Technological Research Council of Turkey (T\"{U}B\.ITAK) under Grant
No. 111T232. The authors would like to thank \.I.~Adagideli,
\"{O}.~Er\c{c}etin, G.~Karpat, C.~Sa\c{c}l{\i}o\u{g}lu, and
L.~Suba\c{s}{\i} for helpful discussions.


\begin{thebibliography}{99}
\bibitem{Herbert} N. Herbert, Found. Phys. \textbf{12}, 1171 (1982).
\bibitem{Wootters-Zurek} W. K. Wootters and W. H. Zurek, Nature (London) \textbf{299}, 802 (1982).
\bibitem{Dieks} D. Dieks, Phys. Lett. \textbf{92A}, 271 (1982).
\bibitem{Buzek-Hillery} V. Bu\v{z}ek and M. Hillery, Phys. Rev. A \textbf{54}, 1844 (1996).
\bibitem{Bruss1} D. Bru{\ss}, D. P. DiVincenzo, A. Ekert, C. A. Fuchs, C. Macchiavello, and J. A. Smolin, Phys. Rev. A \textbf{57}, 2368 (1998).
\bibitem{Gisin-Massar} N. Gisin and S. Massar, Phys. Rev. Lett. \textbf{79}, 2153 (1997).
\bibitem{Werner} R. F. Werner, Phys. Rev. A \textbf{58}, 1827 (1998).
\bibitem{Fan} H. Fan, K. Matsumoto, and M. Wadati, Phys. Rev. A \textbf{64}, 064301 (2001).
\bibitem{Scarani} V. Scarani, S. Iblisdir, and N. Gisin, Rev. Mod. Phys. \textbf{77}, 1225 (2005).
\bibitem{FanR} H. Fan, Y. -N. Wang, L. Jing, J. -D. Yue, H. -D. Shi, Y. -L. Zhang, and L. -Z.
Mu, arXiv:quant-ph/1301.2956, Phys. Rep. (to be published).
\bibitem{Gisin} N. Gisin, Phys. Lett. A \textbf{242}, 1 (1998).
\bibitem{Weinberg1} S. Weinberg, Ann. Phys. (N.Y.) \textbf{194}, 336 (1989).
\bibitem{Weinberg2} S. Weinberg, Phys. Rev. Lett. \textbf{62}, 485 (1989).
\bibitem{Gisin1} N. Gisin, Phys. Lett. A \textbf{143}, 1 (1990).
\bibitem{Czachor} M. Czachor, Found. Phys. Lett. \textbf{4}, 351 (1991).
\bibitem{Polchinski} J. Polchinski, Phys. Rev. Lett. \textbf{66}, 397 (1991).
\bibitem{SimonT} C. Simon, arXiv:quant-ph/0103057v1.
\bibitem{Schrodinger} E. Schr\"{o}dinger, Proc. Camb. Phil. Soc. \textbf{32}, 446 (1936).
\bibitem{Jaynes} E. T. Jaynes, Phys. Rev. \textbf{108}, 171 (1957).
\bibitem{Hadjisavvas} N. Hadjisavvas, Lett. Math. Phys. \textbf{5}, 327 (1981).
\bibitem{Gisin2} N. Gisin, Helv. Phys. Acta \textbf{62}, 363 (1989).
\bibitem{Hughston} L. P. Hughston, R. Jozsa, and W. K. Wootters, Phys. Lett. A \textbf{183}, 14 (1993).
\bibitem{Mermin} N. D. Mermin, Found. Phys. \textbf{29}, 571 (1999).
\bibitem{Kirkpatrick} K. A. Kirkpatrick, Found. Phys. Lett. \textbf{19}, 95 (2006).
\bibitem{Gedik} Z. Gedik, (unpublished).
\end{thebibliography}
\end{document}